\documentclass[aps,12pt]{revtex4}

\usepackage{color,amsmath,amssymb}
\usepackage[pdftex]{graphicx}

\begin{document}

\title{Can car density in a two-lane section
depend on the position of the latter in a single-lane road?}

\author{N. C. Pesheva~$^{\dag^1}$\email{nina@imbm.bas.bg} and J. G. Brankov~$^{\dag^1\dag^2}$
\email{brankov@theor.jinr.ru}}

\affiliation{$^{\dag^1}$~Institute of Mechanics, Bulgarian Academy
of Sciences, 4 Acad. G. Bonchev St., 1113 Sofia, Bulgaria}
\affiliation{$^{\dag^2}$~Bogoliubov Laboratory of Theoretical
Physics, Joint Institute for Nuclear Research, 141980 Dubna, Russian Federation}

\begin{abstract}

We report here results on the study of the totally asymmetric
simple exclusion processes (TASEP), defined on an open network,
consisting of head and tail simple chain segments with a
double-chain section inserted in-between. Results of numerical
simulations for relatively short chains reveal an interesting new
feature of the network. When the current through the system takes
its maximum value, a simple translation of the double-chain
section forward or backward along the network, leads to a sharp
change in the shape of the density profiles in the parallel
chains, thus affecting the total number of cars in that part of
the network. In the symmetric case of equal injection and ejection
rates $\alpha = \beta >1/2$ and equal lengths of the head and tail
sections, the density profiles in the two parallel chains are
almost linear, characteristic for the coexistence line (shock
phase). Upon moving the section forward (backward), their shape
changes to the one typical for the high (low) density phases of a
simple chain. The total bulk density of cars in a section with a
large number of parallel chains is evaluated too. The observed
effect might have interesting implications for the traffic flow
control as well as for biological transport processes in living
cells. An explanation of this phenomenon is offered in terms of
finite-size dependence of the effective injection and ejection
rates at the ends of the double-chain section.

\vspace{7mm}


\end{abstract}

\maketitle

\section{Introduction}

Traffic often takes place along linear tracks which are interconnected to
form a network structure. However, in the presence of hard-core exclusion the collective
behavior of such transportation  networks is not yet well understood. Indeed,
the role of junctions has been
considered only recently on the example of simple networks involving no more
than two junctions, see \cite{BPB, PK05} or on self-similar tree-like
topologies, see \cite{BM10} and references therein.

\begin{figure}
\centering
\includegraphics[width=5in]{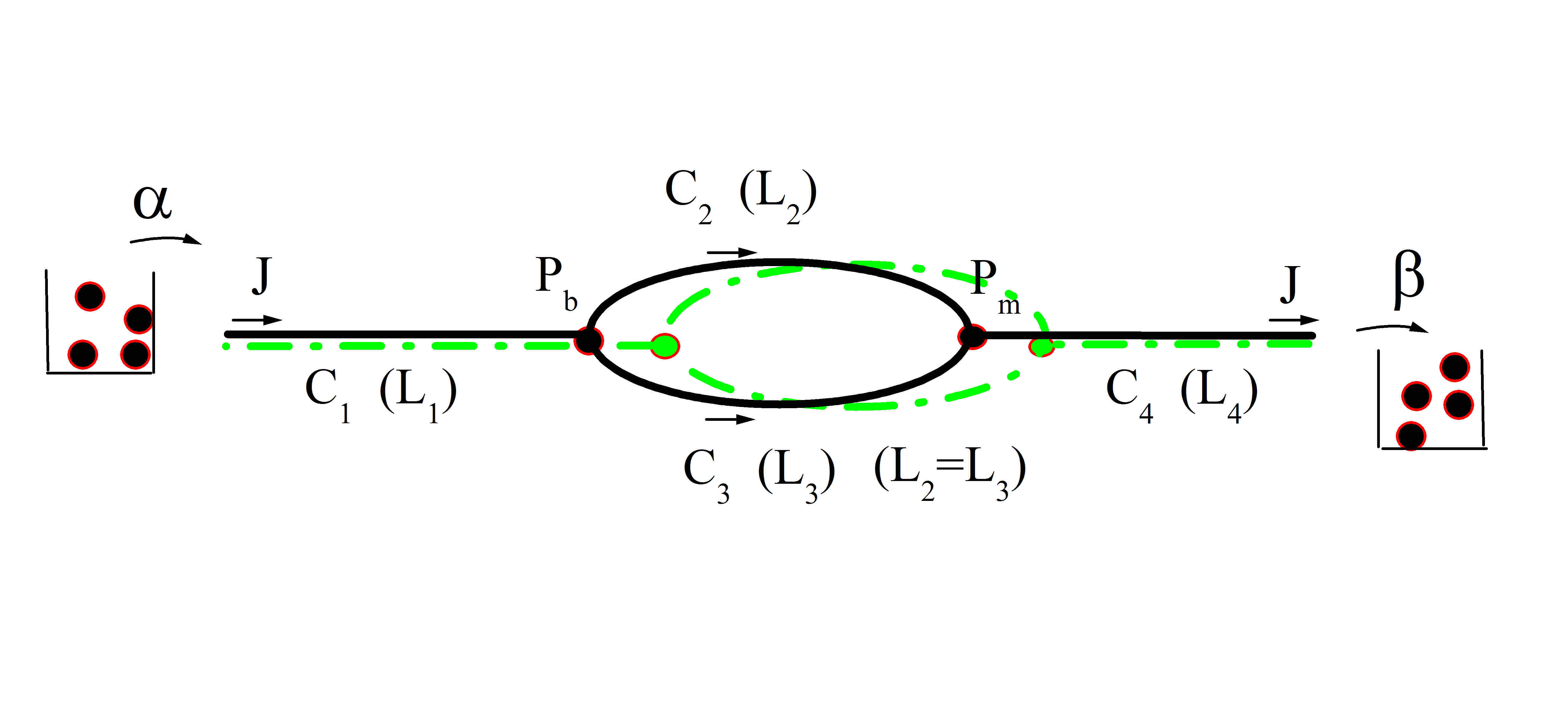}  
\caption{(Color online) Schematic presentation of the considered
system: TASEP, defined on an open network (solid black line),
consisting of head $C_1$ and tail $C_4$ simple chain segments with
a double-chain section ($C_2$ and $C_3$) inserted in-between. In
the system, considered in \cite{BPB} all the simple chain segments
have equal lengths $L_1=L_2=L_3=L_4$ (more details are in the
text). The light grey (green) dash-dotted line shows a version of
the network, where the double-chain section has been translated
forward by $\Delta L$ so that the head and tail simple chain
segments have lengths:  $L_1=L_1+\Delta L$ and $L_4=L_4-\Delta
L$.} \label{Fig1}
\end{figure}

The idea of studying networks composed of chain segments, which
exhibit the bulk behavior of an open TASEP under boundary
conditions, given in terms of effective input and output rates,
was first advanced in our work \cite{BPB}. The network considered
there consists of two vertices of degree 3: one of out-degree 2
and the other one of out-degree 1, connected by 2 chains of the
same direction; the third edge of each of these vertices belongs
to a directed chain coupled to a separate reservoir of particles,
see Fig. \ref{Fig1}. The appearance of correlation effects close
to the ends of the chain segments, as well as of
 cross-correlations in the double-chain segment was demonstrated.
 The same approach was applied in Ref. \cite{PK05} to an open network
  consisting of one vertex of degree 3 and out-degree 1. The two incoming
  chains are coupled to one reservoir, and the outgoing one is coupled
  to another reservoir. Different versions of simple networks were
  studied in Refs. \cite{EPK08} and \cite{EPK09}. In the former
   reference two cases were investigated: (a) two vertices of
   degree 3 connected by three chains, one of which has the opposite
    direction to the remaining two (closed system); (b) two vertices
    of degree 3 connected by two chains with the same direction;
    the remaining incoming and outgoing chains are coupled to reservoirs
     with the same particle density (open counterpart). In Ref. \cite{EPK09}
     graphs containing vertices of degree 4 and out-degree 1, 2, and 3 were considered.
     The notion of particle-hole symmetry in the presence of a junction was
     carefully analyzed and an appropriate interpretation on the microscopic level
     was given. TASEP with parallel update on single multiple-input--single-output
     junctions has been investigated too \cite{WLJ08}. Clearly, the above works
     have treated TASEP on diverse, but simple fixed network topologies.
     The main concern was the construction of the phase diagram under
     different open boundary conditions.

Complex networks have also been a focus of research in the last decade.
 Attention has been paid to the traffic fluctuation problem in networks:
 the dependence between the mean value of the of traffic passing through
 a node (or a link) in a time interval and its standard deviation has
 been studied. It has been recognized that in the evolving networks
 functional units may emerge, which may be represented by topologically
 distinct subgraphs, see \cite{KN09} and references therein. Different
  types of diffusive dynamics, like spreading of disease and traffic or
  navigated walks have been  studied on different networks, see, e.g.,
  \cite{T01, Getal02, NR04}. In particular, high-density traffic of
  information packets on sparse modular networks with scale-free
   subgraphs was studied, see \cite{TM09}. Most of the results of graph
    theory relevant to large complex networks were related to the simplest
    models of random graphs. A large amount of the research was devoted
    to scale-free networks, i.e., networks with power-law vertex degree
     distribution. Traffic rules on such graphs usually include particle
     creation at randomly selected nodes and  mutually interacting random
     walks to different specified destinations, see \cite{Eetal05}.
In \cite{NKP11} the authors studied the stationary transport properties
of the TASEP with random-sequential update on complex networks, both
deterministic and stochastic. The TASEP rules were generalized, by an
 obvious extension of the rules applied in \cite{BPB}, to fixed connected
 networks of $N_s$ directed segments (each consisting of $L\gg 1$ sites)
  and $N_v$ nodes. The theoretical approach was based on a combination
  of models of complex networks with the well-known mean field (MF) results
   for the simple chains ('segments') which are assumed to connect the vertices
   of the underlying directed graph. These chain segments, representing the
    edges of the network graph, have to be long enough to make reasonable
     the application of the MF results. At that, the correlations that may
      build up close to the nodes of the network have to be neglected.

Recently, applications to biological transport have motivated generalizations
of the TASEP to cases when the entry rate is chosen
to depend on the number of particles in the reservoir (TASEP with finite resources)
 \cite{CZ09}. Last year, the cases of multiple competing TASEPs
with a shared reservoir of particles \cite{GCAR12} and with limited
 reservoirs of particles and fuel carriers \cite{BCR12} were studied too.

\section{Model and numerical simulation results}

In the asymmetric simple exclusion process, hard-core particles
move along one-dimensional lattice of sites, which can be occupied
by one particle at most $\{\tau_i=0,1\}, \, i=1,\dots, L$. The
particle can move right (left) with probability $p$ $(1-p)$ to a
nearest neighbor site, if the site is empty. In the extremely
asymmetric case particles are allowed to move in one direction
only --- this is the totally asymmetric simple exclusion process
(TASEP). Its steady states are exactly known for both open and
periodic boundary conditions, for continuous-time and several
kinds of discrete-time dynamics. Here, we focus our attention on
the steady states of the open TASEP with continuous-time
stochastic dynamics, modeled by the so called random-sequential
update. For a review on the exact results for the stationary
states of TASEP,  under different kinds of stochastic dynamics,
and its numerous applications, we refer the reader to \cite{RSSS,
ERS, S01}.

   Our goal here is to present some interesting new effects,
observed in a TASEP, defined on a simple network, consisting of
head and tail simple chain segments ($C_1$ and $C_4$) with a
double-chain section ($C_2$ and $C_3$) inserted in-between. The
model system is presented schematically in  Fig. \ref{Fig1}.
 The particle injection rate at the left end of the network is $\alpha$ and the particle
ejection rate at the right end of the network is $\beta$.
 $P_b$ is the branching point --- the last site $(i=L_1)$ of the head chain segment
 $C_1$, where particles can take with equal probability $p_j$ the upper $C_2$
  or the lower $C_3$ branch of the double-chain section. $P_m$
  is the merging point (first site of the tail segment $i=L_1+L_2+1$), where the
 particles moving along  the $C_2$ and $C_3$ chain segments  merge. The particle
current in the system is denoted by $J$ (it is in the same
direction for both chains of the double-chain section).

\begin{figure}
\centering
\includegraphics[width=4in]{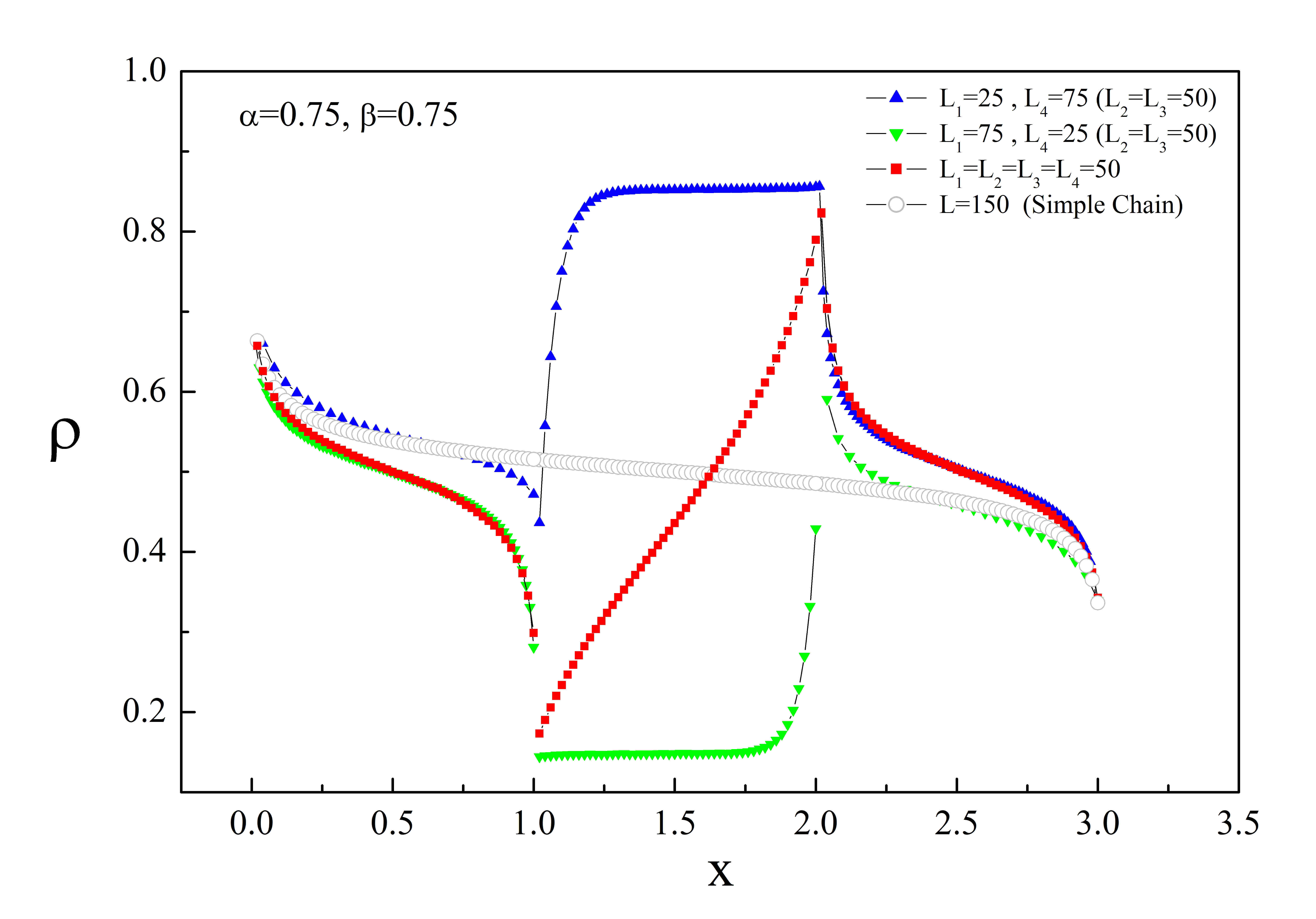}  
\caption{(Color online) Density distribution $\rho (x)$ (from
numerical simulations) along the different chain segments of the
open network of the type, shown in Fig. 1, when  $\alpha=0.75$,
$\beta=0.75$. The solid (red) squares  are the reference results
for the network with the  (MC,CL,MC) phase structure appearing
when $L_1=L_2=L_3=L_4$. Simple translation of  the double-chain
section, while the lengths of the double chain segments $C_2$ and
$C_3$ are kept fixed $L_2=L_3=50$, and also the length of the
whole network is kept fixed at $L=150$ sites, causes a noticeable
change of the density profiles of $C_2$ and $C_3$. Translation
 forward leads to a density distribution shape on $C_2$ and $C_3$
 (shown with  solid (green) down-triangles), characteristic
  of the LD phase, while translation backward to a density
  distribution (solid (blue) up-triangles), characteristic of the HD phase.
For comparison the density distribution (in a MC phase) of a
simple chain of length $150$ sites is also shown with empty grey
circles of larger size. } \label{Fig2}
\end{figure}

\begin{figure}
\centering
\includegraphics[width=4in]{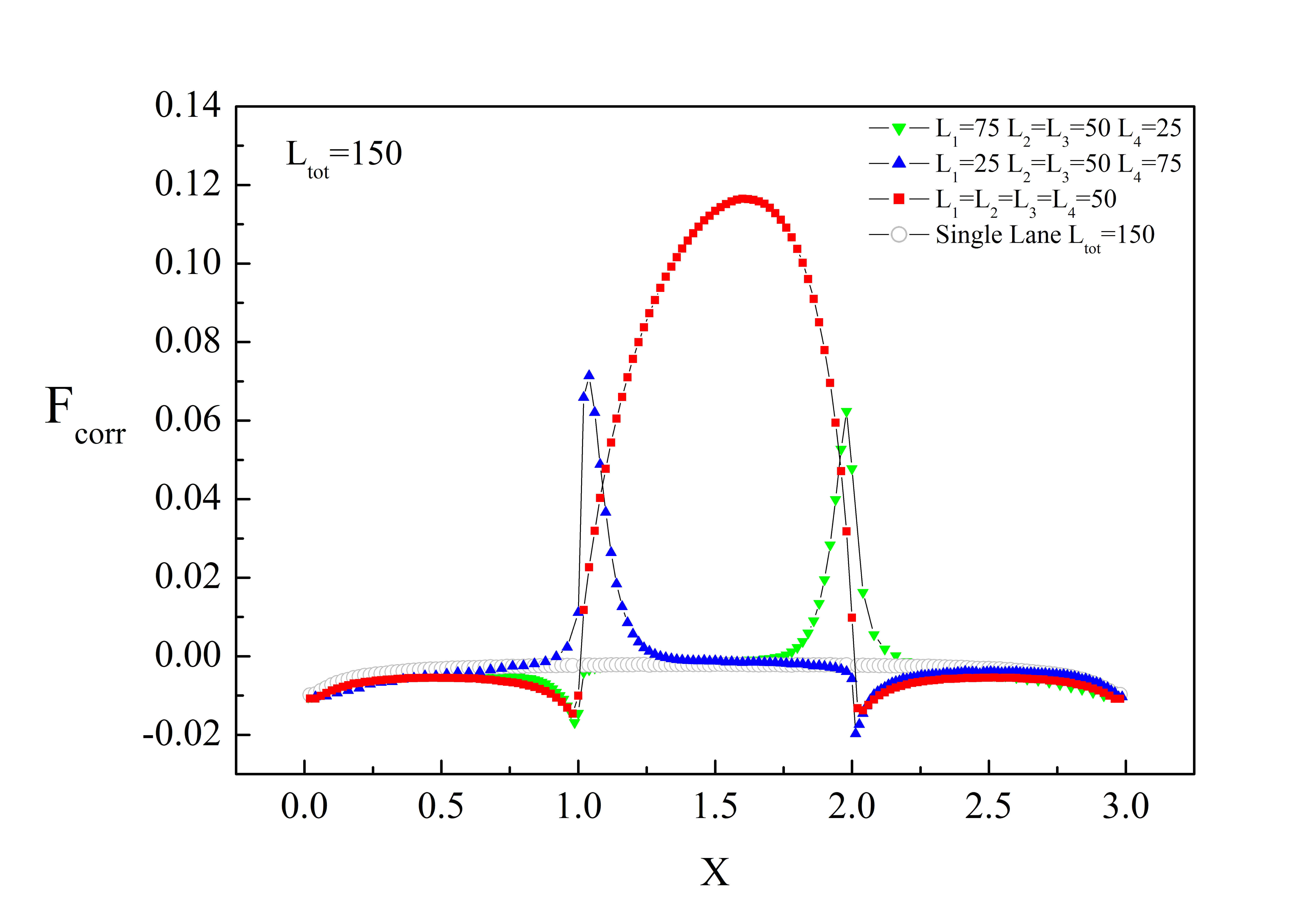}  
\caption{(Color online) Nearest-neighbor correlations $F_{corr}$ in
the network with a double-chain section  as a function of the
scaled distance
$x$ 
for $\alpha=0.75$, $\beta=0.75$.  The simulation
results for $F_{corr}$ correspond to the density distributions,
shown in Fig. 2, for the different positions of the double-chain
section. } \label{Fig3}
\end{figure}

\begin{figure}
\centering
\includegraphics[width=4in]{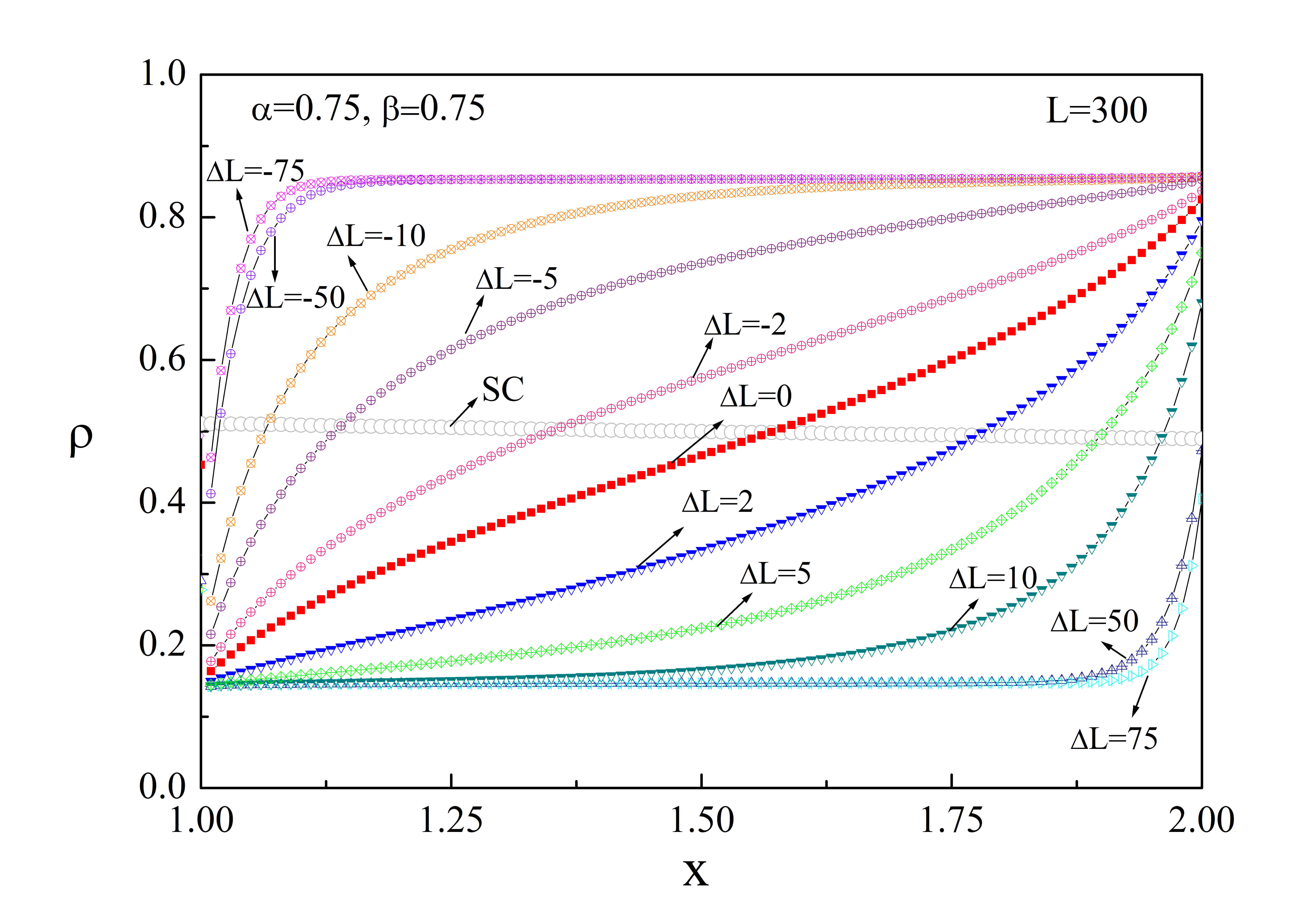}  
\caption{ (Color online) Density distribution $\rho (x)$
(simulation results) along the two equivalent chain segments of
the double chain section for bigger size network (total length
$L=300$ sites) at $\alpha=0.75$, $\beta=0.75$. The results show
the increased sensitivity of the density distributions to even a
very small position change $\Delta L$ of the double chain section.
The solid (red) squares are the reference results for
$L_1=L_2=L_3=L_4=100$ ($\Delta L$=0).} \label{Fig4}
\end{figure}

 The phase structure of the network, studied in \cite{BPB} when all the simple
chain segments have equal length, $L_1=L_2=L_3=L_4$, is presented
by the triplet $(X_1,X_{2,3},X_4)$, where $X_n\  (n = 1,2,3,4)$
stands for one of the stationary phases of the simple chain
segment $C_n$: LD --- low density, HD --- high density, MC ---
maximum current, and CL --- coexistence line. Our analytical
analysis of the allowed phase structures, based on the properties
of single chains in the thermodynamic limit, and the neglect of
the pair correlations between the nearest-neighbor occupation
numbers at the junctions of the different chain segments, yielded
8 possibilities. Here we focus our investigation on 3 of the most
interesting cases (MC,LD,MC), (MC,CL,MC), and (MC,HD,MC) which
appear when the boundary rates satisfy the inequalities $\alpha>
1/2, \ \beta > 1/2$, corresponding to the maximum current phase of
a single chain. We have shown that the phase state of the chains
in the double-chain section depends on the effective injection
rate $\alpha^*$ of particles at the first site of each of the
chain segments $C_{2,3}$ and on the effective removal rate
$\beta^*$ of particles from the last site of each of these chains.
Our further studies of the system reveal a rather interesting
property whenever the network with $L_1=L_2=L_3=L_4$ has the
(MC,CL,MC) phase structure.
 One can change effectively the
effective rates $\alpha^*$ and $\beta^*$ at the network junction
points simply by changing the position of the double chain section
along the network, while keeping fixed both: the total length $L$
of the network and the length $L_2=L_3=L_d$ of the double-chain
section. This is clearly observable in Figs.
\ref{Fig2}--\ref{Fig4}. As one can see in Fig. \ref{Fig2}, a
simple translation forward of the double chain section leads to a
density profile on $C_2$ and $C_3$ characteristic of the LD phase
(shown with  solid down-triangles), while translation backward
induces a density profile characteristic of the HD phase (shown
with solid up-triangles). As a reference, the results for the
density profiles of the system with segments of equal length
$L_1=L_{2,3}=L_4=50$ are shown with solid  squares.

The characteristic change of the density profiles is accompanied
also with the corresponding characteristic change in the
nearest-neighbor correlation function $F_{corr}$, displayed in
Fig. \ref{Fig3}, for the same cases, as the ones shown in Fig.
\ref{Fig2}. For comparison the density distribution (in the MC
phase) and nearest-neighbor correlations of a simple chain of
length $L=150$ sites are also shown in both  figures with empty
grey circles of larger size.
 The Monte Carlo simulation results, presented here,
are for a relatively small system of fixed total length $L = L_1 +
L_{2,3} + L_4 = 150$ sites and fixed size of the double-chain
section, $L_2 = L_3 = 50$. The ensemble averaging was performed
over $200$ independent runs and $1.5\times 10^6$ Monte Carlo steps
were omitted in order to ensure that the system has reached a
stationary state.

Numerical study of a larger system with total length
$L=300$ sites reveals even higher sensitivity of the density
distribution $\rho(x)$ along the chains $C_2$ and $C_3$ of the
double-chain section with respect to quite small changes of the
loop position on the network, see Fig. \ref{Fig4}. Already at $\Delta L=2$ one can
observe a noticeable change in the density distribution.

\section{Theoretical analysis}

\subsection{Mean field theory}

The possible phase structure of the present network was analyzed
in our paper \cite{BPB} by using the exact results for simple bulk
chains $C_n$, $n = 1,2,3,4$, coupled to each other by means of
effective injection and removal rates. To be specific, let
$\tau_i^{(n)}$, $i=1,2,\dots ,L_n$, denote the particle occupation
number of site $i$ in chain $C_n$ which contains $L_n$ sites. Here
we let the length of the double-chain defect $L_2 =L_3\equiv L_d$
and the total length of the network $L_1+ L_d + L_4 \equiv L$ to
be fixed. The mean field theory assumes that all $L_n$, $n =
1,2,3,4$, are large enough, so that the average occupation number
$\langle\tau_i^{(n)}\rangle(L;\alpha, \beta)$ of site $i\in C_n$
in the stationary state of the whole system, under injection rate
$\alpha$ and removal rate $\beta$, can be identified with the
average occupation number
$\langle\tau_i^{(n)}\rangle(L_n;\alpha^{*}, \beta^{*})$ in the
stationary state of a single chain of length $L_n$ under some
effective injection ($\alpha_n^{*}$) and removal $(\beta_n^{*})$
rates. Note that at the open boundaries of the whole network the
continuity of the current $J_L$ through the system implies
\begin{equation}
J_L= \alpha [1- \langle\tau_1^{(1)}\rangle(L_1;\alpha, \beta)] =
\beta \langle\tau_L^{(4)}\rangle(L_4;\alpha, \beta).
\label{J1}
\end{equation}
Under the above assumptions, these equalities will be approximated by
\begin{equation}
J_L\simeq \alpha [1- \langle\tau_1^{(1)}\rangle(L_1;\alpha, \beta_1^{*})] \simeq
\beta \langle\tau_{L_4}^{(4)}\rangle(L_4;\alpha_4^{*},\beta).
\label{J2}
\end{equation}
Similarly, by neglecting the nearest-neighbor correlations at the junctions of different
chains, the current continuity equation yields the approximate equalities
\begin{eqnarray}
&&J_L\simeq  \langle\tau_{L_1}^{(1)}\rangle(L_1;\alpha, \beta_1^{*})\left\{\frac{1}{2}\left[1- \langle\tau_1^{(2)}\rangle(L_d;\alpha_2^{*}, \beta_2^{*})\right] +\frac{1}{2}
\left[1- \langle\tau_1^{(3)}\rangle(L_d;\alpha_3^{*}, \beta_3^{*})\right]\right\} \nonumber \\ &&\simeq
\left[\langle\tau_{L_d}^{(2)}\rangle(L_d;\alpha_2^{*}, \beta_2^{*})+\langle\tau_{L_d}^{(3)}\rangle(L_d;\alpha_3^{*}, \beta_3^{*})\right]\left[1- \langle\tau_1^{(4)}\rangle(L_4;\alpha_4^{*},\beta)\right].
\label{J3}
\end{eqnarray}
Hence, taking into account the equivalence of chains $C_2$ and $C_3$, we set $\alpha_{2}^{*}=\alpha_{3}^{*}=\alpha_{d}^{*}$,
$\beta_{2}^{*}=\beta_{3}^{*}=\beta_{d}^{*}$, and define
\begin{eqnarray}
&& \beta_1^{*} = 1- \langle\tau_1^{(2,3)}\rangle(L_d;\alpha_{d}^{*}, \beta_{d}^{*}), \qquad
\beta_{d}^{*} = 1- \langle\tau_1^{(4)}\rangle(L_4;\alpha_{4}^{*}, \beta) \nonumber \\
&&\alpha_{d}^{*} =\frac{1}{2}\langle\tau_{L_1}^{(1)}\rangle(L_1;\alpha, \beta_1^{*}), \qquad
\alpha_4^{*} = 2\langle\tau_{L_d}^{(2,3)}\rangle(L_d;\alpha_{d}^{*}, \beta_{d}^{*}),
\label{eff}
\end{eqnarray}
or, alternatively,
\begin{eqnarray}
&& \beta_1^{*} =\frac{J_L}{\langle\tau_{L_1}^{(1)}\rangle(L_1;\alpha, \beta_1^{*})}, \qquad
\beta_{d}^{*} = \frac{J_L}{2\langle\tau_{L_d}^{(2,3)}\rangle(L_d;\alpha_{d}^{*}, \beta_d^*)} \nonumber \\
&&\alpha_{d}^{*} =\frac{J_L}{2[1- \langle\tau_{1}^{(2,3)}\rangle(L_d;\alpha_{d}^{*}, \beta_d^*)]}, \qquad
\alpha_4^{*} = \frac{J_L}{1- \langle\tau_{1}^{(4)}\rangle(L_4;\alpha_{4}^{*}, \beta)}.
\label{eff2}
\end{eqnarray}

Thus, definitions (\ref{eff}) express the effective injection (removal) rates of chains $C_{2,3}$, $C_4$
($C_1$, $C_{2,3}$) in terms of the average occupation number of the last (first) site of the preceding
(subsequent) chain, while definitions (\ref{eff2}) express the corresponding effective injection (removal)
rates in terms of the finite-size current and the average occupation number of the first (last) site of the same
chain. The consistency of these definitions is a measure of the extent to which nearest-neighbor correlations
at the junctions can be neglected.

The above equations essentially simplify when $L_n \gg 1$, $n=1,2,3,4$, by replacing the finite-size properties of the current and the density profiles with their thermodynamic counterparts. To this end we define
\begin{eqnarray}
&&\rho_{in}^{(n)}(\alpha,\beta)= \lim_{L_n \rightarrow \infty} \langle\tau_1^{(n)}\rangle(L_n;\alpha,\beta), \quad
\rho_{out}^{(n)}(\alpha,\beta)= \lim_{L_n \rightarrow \infty} \langle\tau_{L_n}^{(n)}\rangle(L_n;\alpha,\beta), \nonumber \\ &&
\rho_{bulk}^{(n)}(\alpha,\beta)= \lim_{L_n \rightarrow \infty}\frac{1}{L_n}\sum_{i=1}^{L_n} \langle\tau_i^{(n)}\rangle(L_n;\alpha,\beta).
\label{ro}
\end{eqnarray}
In this case the current is determined by the bulk densities (omitting the dependence
on the injection and removal rates):
\begin{equation}
J = \lim_{L \rightarrow \infty}J_L = \rho_{bulk}^{(1,4)}(1- \rho_{bulk}^{(1,4)}) = 2\rho_{bulk}^{(2,3)}(1- \rho_{bulk}^{(2,3)}).
\end{equation}
Hence, we find expressions for the possible bulk densities in terms of the current in the thermodynamic limit:
\begin{equation}
\rho_{bulk}^{(1)}=\rho_{\pm}(J),\qquad  \rho_{bulk}^{(4)}=\rho_{\pm}(J),
\qquad  \rho_{bulk}^{(2,3)} = \rho_{\pm}(J/2),
\label{robulk}
\end{equation}
where
\begin{equation}
\rho_{\pm}(J) = \frac{1}{2}\left(1\pm \sqrt{1-4J}\right).
\label{ropm}
\end{equation}

In the most interesting case $\alpha > 1/2$ and $\beta >1/2$, both chains $C_1$ and $C_4$
are in the maximum current (MC) phase \cite{BPB}, when the exact results in
the limit $L\rightarrow \infty$
yield
\begin{eqnarray}
&& J=1/4, \quad \rho_{bulk}^{(1)}(\mathrm{MC})= \rho_{bulk}^{(4)}(\mathrm{MC}) = 1/2,
\nonumber \\ &&\rho_{in}^{(1)}(\alpha, \beta_1^{*}) = 1-\frac{1}{4\alpha},
\quad \rho_{out}^{(1)}(\alpha, \beta_1^{*}) = \frac{1}{4\beta_1^{*}},
\nonumber \\
&& \rho_{in}^{(4)}(\alpha_{4}^{*}, \beta)= 1-\frac{1}{4\alpha_{4}^{*}}, \quad
\rho_{out}^{(4)}(\alpha_4^{*},\beta)= \frac{1}{4\beta}.
\label{MCbulk}
\end{eqnarray}
In this case from Eq. (\ref{robulk}) it follows that either $\rho_{bulk}^{(2,3)} = \rho_{-}(1/8)  \simeq 0.146$
and the defect chains $C_{2,3}$ are in the low density (LD) phase, or $\rho_{bulk}^{(2,3)} = \rho_{+}(1/8) \simeq 0.854$
and the defect chains $C_{2,3}$ are in the high density (HD) phase.

Consider first the LD case, when
\begin{equation}
\rho_{in}^{(2,3)}(\alpha_d^{*}, \beta_d^{*})= \rho_{bulk}^{(2,3)}(\alpha_d^{*}, \beta_d^{*})=\alpha_d^{*},\quad
\rho_{out}^{(2,3)}(\alpha_d^{*}, \beta_d^{*})= \alpha_d^{*}(1-\alpha_d^{*})\beta_d^{*}.
\label{LDC23}
\end{equation}
Then, in the large $L$ limit Eqs. (\ref{eff}) and  (\ref{eff2}) yield
\begin{equation}
\alpha_{d}^{*} =\rho_{bulk}^{(2,3)}(\alpha_d^{*}, \beta_d^{*}) =\rho_{-}(1/8) , \quad \beta_{1}^{*} = \frac{1}{8\alpha_{d}^{*}}= \rho_{+}(1/8) ,
\quad \beta_{d}^{*} =  \frac{1}{4\alpha_{4}^{*}} =\frac{1}{8\rho_{out}^{(2,3)}(\alpha_{d}^{*}, \beta_{d}^{*})}.
\label{LD}
\end{equation}

Consider next the HD case, when
\begin{equation}
\rho_{in}^{(2,3)}(\alpha_d^{*}, \beta_d^{*})=  1-\beta_d^{*}(1-\beta_d^{*})/ \alpha_d^{*},     \quad
\rho_{out}^{(2,3)}(\alpha_d^{*}, \beta_d^{*})= \rho_{bulk}^{(2,3)}(\alpha_d^{*}, \beta_d^{*})= 1- \beta_d^{*}.
\label{HDC23}
\end{equation}
Then,  the large $L$ limit of  Eqs. (\ref{eff}) and  (\ref{eff2}) yields
\begin{equation}
\alpha_{d}^{*} =\frac{1}{2}\rho_{out}^{(1)}(\alpha, \beta_{1}^{*}) =\frac{1}{8\beta_{1}^{*}},
\quad \alpha_{4}^{*} = \frac{1}{4\beta_{d}^{*}}=2\rho_{+}(1/8),
\quad \beta_{d}^{*} =1 - \rho_{bulk}^{(2,3)}(\alpha_d^{*}, \beta_d^{*})  = \rho_{-}(1/8) .
\label{HD}
\end{equation}

We see that in both cases there remain undetermined effective rates: in the low density case
these are $\beta_{d}^{*}$ and the related  $\alpha_{4}^{*}$, see Eq. (\ref{LD}), and
in the high density case  $\alpha_{d}^{*}$ and the related $\beta_{1}^{*}$, see Eq. (\ref{HD}).
We explain this feature by the fact that the relation between the magnitudes of $\alpha_{d}^{*}$
and $\beta_{d}^{*}$ depends on the phase of the chains $C_{2,3}$ when chains $C_1$ and $C_4$ are
in the maximum current phase. Then the current through the whole system attains its maximum
value $J =1/4$ and the bulk densities in $C_1$ and $C_4$ take the only possible value $\rho_{bulk}^{(1)} =
\rho_{bulk}^{(4)} = 1/2$. However, the value of the current through each of the chains $C_{2,3}$,
$J^(2,3) =J/2= 1/8$ allows them to be either in the low density
phase, when $\beta_{d}^{*} > \alpha_{d}^{*}= \rho_{-}(1/8)$, or in the high density phase, when
$\alpha_{d}^{*} > \beta_{d}^{*} = \rho_{-}(1/8)$, or on the coexistence line, when $\alpha_{d}^{*} =
\beta_{d}^{*} = \rho_{-}(1/8)$. This means that there is a degree of freedom due to the fact that
the average density of particles in the chains $C_{2,3}$ is not fixed. In the next section we show
how one can deduce the missing injection/ejection rate from the  fit of the density profile with the
exponential distribution predicted by the domain wall theory.

Our former computer simulations \cite{BPB} have shown that in the
symmetric case of $L_1 = L_4 =200$ and $\alpha = \beta > 1/2$ the
double-chain segment is found to be on the coexistence line
$\alpha_{d}^{*}= \beta_{d}^{*} <1/2$ (known also as a shock
phase) when a completely delocalized domain wall exists. In this
case the local density profile is linear and changes in the
interval from $\rho_{bulk}(\mathrm{LD})$ to
$\rho_{bulk}(\mathrm{HD})$, so that $\rho_d(\alpha_{d}^{*},
\alpha_{d}^{*})=1/2$. Evidently, for a finite-size system the
effective rates depend on the lengths of all the chains, $L_1$,
$L_d$ and $L_4$. Simple arguments lead to the conclusion that as
$L_1 \rightarrow 1$, the effective injection rate $\alpha_{d}^{*}$
monotonically increases, so that $\alpha_{d}^{*} \rightarrow
\alpha/2 > 1/4$, while $\beta_{d}^{*}=\rho_{-}(1/8) <1/4$ does not
change significantly if $L_d \gg 1$, the defect chains $C_{2,3}$
go into the high density phase. On the other hand, when $L_4
\rightarrow 1$, $\beta_{d}^{*}$ monotonically increases, so that
$\beta_{d}^{*} \rightarrow \beta > 1/2$, while
$\alpha_{d}^{*}=\rho_{-}(1/8) <1/4$ does not change significantly,
the defect chains $C_{2,3}$ go into the low density phase. That is
actually observed in our present computer simulations, the results
of which are illustrated in Figs. \ref{Fig2}. The noticeable
deviations of the density profiles from the standard infinite
chain counterparts are due to finite-size effects,  as well as to
the correlations that appear near the points of inhomogeneity of
the network, see Fig. \ref{Fig3}. The sensitivity of the effect
increases with the length of the chains, see Fig. \ref{Fig4},
since then the smeared phase transition in $C_{2,3}$  becomes
sharper.

Some numerical data for a network with fixed $L= 150$, $L_d =50$ and changing $L_1 = 100 - L_4$, are given in Table 1.
The values of $\alpha_{d}^{*}$ and $\beta_{d}^{*}$ are calculated
from the finite-size numerical data for the local densities $\langle\tau_{L_1}^{(1)}\rangle$, $\langle\tau_1^{(4)}\rangle$,
$\langle\tau_1^{(2,3)}\rangle$ and $\langle\tau_{L_d}^{(2,3)}\rangle$,  by using equations (\ref{eff}) and (\ref{eff2})
with $J_L = J(\mathrm{MC}) = 1/4$.
\vspace{1cm}
\begin{center}
\begin{tabular}{|c|c|c|c|c|c|c|c|c|}  \hline
$\; L_1 \;$ & \; $\langle\tau_{L_1}^{(1)}\rangle$ & \; $\langle\tau_1^{(4)}\rangle$
& $\langle\tau_1^{(2,3)}\rangle$ & $\langle\tau_{L_d}^{(2,3)}\rangle$ & $\alpha_{d}^{*}$ & $\alpha_{d}^{*}$&
$\beta_{d}^{*}$ & $\beta_{d}^{*}$
\\ & & & & & from Eq. (\ref{eff})& from Eq. (\ref{eff2}) & from Eq. (\ref{eff})& from Eq. (\ref{eff2}) \\
 \hline 25 &  0.472 & 0.856 & 0.436 & 0.856 & 0.236 & 0.222 & 0.144 & 0.146
\\ \hline 50 &  0.299 & 0.824 & 0.175 & 0.791 & 0.150 & 0.152 & 0.176 & 0.158
\\ \hline 75 & 0.281 & 0.590 & 0.144& 0.429 & 0.140 & 0.146 & 0.410 & 0.291 \\ \hline
\end{tabular}
\\
\vspace{0.5cm}
Table 1.
\end{center}
\vspace{1cm}

As is seen, the relative values of the effective rates change with
the defect position, according to our expectations, and properly
describe the phase changes, observed in the density profiles.
 The largest numerical discrepancies are
observed between the values of $\beta_{d}^{*}$ obtained from Eqs.
(\ref{eff}) and  (\ref{eff2})  in the cases of networks with $L_1
= 50$ and $L_1 = 75$. Probably, they can be explained by taking
into account the high nearest-neighbor correlations between the
defect section and the tail chain.

\subsection{Domain wall theory}

According to the Domain Wall (DW) theory, the configurations of TASEP
on a simple chain with open boundaries can be approximated by
two regions, of low and high density, separated by a domain law of
zero width \cite{KSKS}, see also \cite{SA}. Thus, all the configurations
of a chain of $L$ sites
can be labeled by a single integer $k=0,1,\dots, L$, such that sites
$0\leq i \leq k$ belong to the low density phase with uniform density
$\rho_-$, and sites  $k+1 \leq i \leq L$ belong to the high density phase
with uniform density $\rho_+$. The extremal values of $k=0$ and $k=L$
are understood to label configurations corresponding to the pure high
and low density phases.

We recall that the probability $P(k,t)$ for finding at time $t$ the domain wall at position $k$
satisfies a master equation with reflecting boundary conditions at $k=0$ and $k=L$.
The corresponding stationary solution $P_*(k)$ has the simple exponential form
\begin{equation}
P_*(k) = r^{-k} /{\mathcal N},
\label{Pst}
\end{equation}
where
\begin{equation}
r = \frac{\alpha(1-\alpha)}{\beta(1-\beta)} = \left\{ \begin{array}{cc} \exp(-1/\xi),& \alpha <\beta \leq 1/2 \nonumber \\
\exp(1/\xi), & \beta<\alpha \leq 1/2 \end{array}\right.,
\label{rdef}
\end{equation}
where $\xi >0 $ is the domain wall localization length; the
normalization factor ${\mathcal N}$ is
\begin{equation}
{\mathcal N} := \sum_{k=0}^L r^{-k} = \frac{1- r^{-(L+1)}}{1- r^{-1}}.
\label{Ndef}
\end{equation}

Thus, the local density of particles $\rho(i) := \langle n_i \rangle$ is given by
\begin{equation}
\rho(i)= \rho_+ {\mathcal N}^{-1} \sum_{k=0}^{i-1} r^{-k} + \rho_- {\mathcal N}^{-1}
\sum_{k=i}^{L} r^{-k} = \rho_- + (\rho_{+} - \rho_{-})\frac{1- r^{-i}}{1- r^{-L-1}}.
\label{roi}
\end{equation}
This expression describes different shapes depending on the value of $r$.

(a) On the coexistence line (CL) $r =1$ and Eq. (\ref{roi}) yields the linear profile
\begin{equation}
\rho_{CL}(i)=  \rho_{-} + (\rho_{+} - \rho_{-})\frac{i}{L+1}, \qquad 1\leq i \leq L.
\label{CLprofF}
\end{equation}

(b) In the low density (LD) phase $r<1$ and up to exponentially small corrections one obtains
\begin{equation}
\rho_{LD}(i)= \rho_{-} + (\rho_{+} - \rho_{-})r^{L-i+1}, \qquad r < 1, \qquad 1\leq i \leq L.
\label{LDprofF}
\end{equation}

(c) In the high density phase $r>1$ and up to exponentially small corrections one obtains
\begin{equation}
\rho_{HD}(i)= \rho_{+} - (\rho_{+} - \rho_{-})r^{-i}, \qquad r >1, \qquad 1\leq i \leq L.
\label{HDprofF}
\end{equation}

We shall concentrate on the last two cases which will be considered on the macroscopic scale $i/L =x$.
To this end it is convenient to introduce the parameter $t = \xi/L$ and rewrite expression  (\ref{LDprofF}) as
\begin{equation}
\tilde{\rho}_{LD}(x)= \rho_{-}  + A\exp{[(x-1-1/L)/t]}, \qquad L^{-1}\leq x \leq 1,
\label{LDpro}
\end{equation}
where, $\tilde{\rho}_{LD}(x) = \rho_{LD}(xL)$ and $A = \rho_{+} - \rho_{-}$.
Similarly, expression  (\ref{HDprofF}) takes the form
\begin{equation}
\tilde{\rho}_{HD}(x)= \rho_{+} - A\exp{(-x/t)}, \qquad L^{-1}\leq x \leq 1,
\label{HDpro}
\end{equation}
where, $\tilde{\rho}_{HD}(x) = \rho_{HD}(xL)$ and $A = \rho_{+} - \rho_{-}$.

The results from the interpretation of our numerical results on the local density
profiles within the DW theory are explained in the following two subsections.

\subsubsection{Interpretation of the density profile fit in the LD phase}

\begin{figure}
\begin{center}
  \includegraphics[width=100mm]{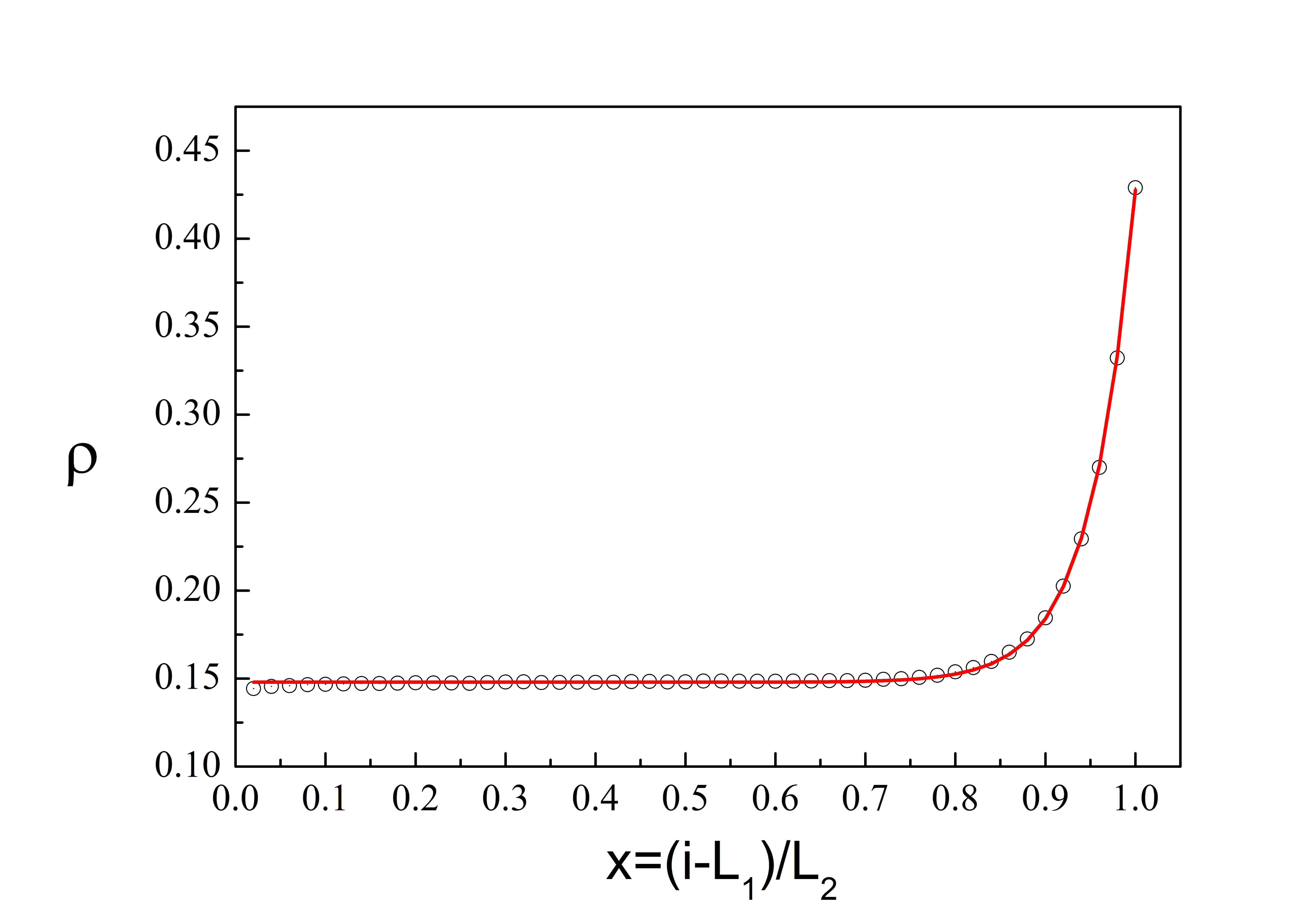}
    \caption{(Color online) The simulation results for the particle
     density profile $\rho(x)$, $x=(i-L_1)/L_2$, $i=L_1 +1,L_1 + 2,\dots, L_1 +L_2$, in the
LD phase of a chain $C_{2,3}$ in the network, are shown by
centered grey circles,
  the fit with the function (\ref{LDpro}) is shown by  solid (red) line.
  The quality of the fit is characterized by the statistical criteria
  $\chi^2/DoF = 1.0004\times 10^{-6}$ and $R^2 = 0.99963$.\label{LDprof}}
  \end{center}
\end{figure}

The fit to the simulated local density profile by function (\ref{LDpro}) is shown in Fig. \ref{LDprof}.
Its  very high quality yields reliable estimates of $\rho_{-} = \alpha_d^* \simeq 0.148$, $A = 0.421 \pm 0.002$
and $t = 0.0488 \pm 0.0003$, hence, $r = \exp(-1/\xi) \simeq 0.664$.

The value of $\alpha_d^* = 0.148$ is in fairly good agreement with
the average current $J^{(2,3)}= J^{max}/2$ through each of the
chains $C_{2,3}$: $\alpha_d^*(1-\alpha_d^*) \simeq 0.126$; the
difference between 0.126 and the expected value $J^{max}/2=1/8$
through each of the chains of the double-chain section can be
attributed to  finite-size effects in a network of total length
$L=150$.

The most important observation is that, by using the prediction of the DW theory,
\begin{equation}
\beta_d^*(1-\beta_d^*) = \alpha_d^*(1-\alpha_d^*)/r =J^{(2,3)}/r,
\end{equation}
we can evaluate the effective ejection rate $\beta_d^*$ by solving the above equation at
$r = 0.664$, with  the result
\begin{equation}
\beta_d^* = (1/2)[1-\sqrt{1 -4J^{(2,3)}/r}] \simeq 0.255.
\label{betaLD}
\end{equation}

\begin{figure}
\begin{center}
  \includegraphics[width=100mm]{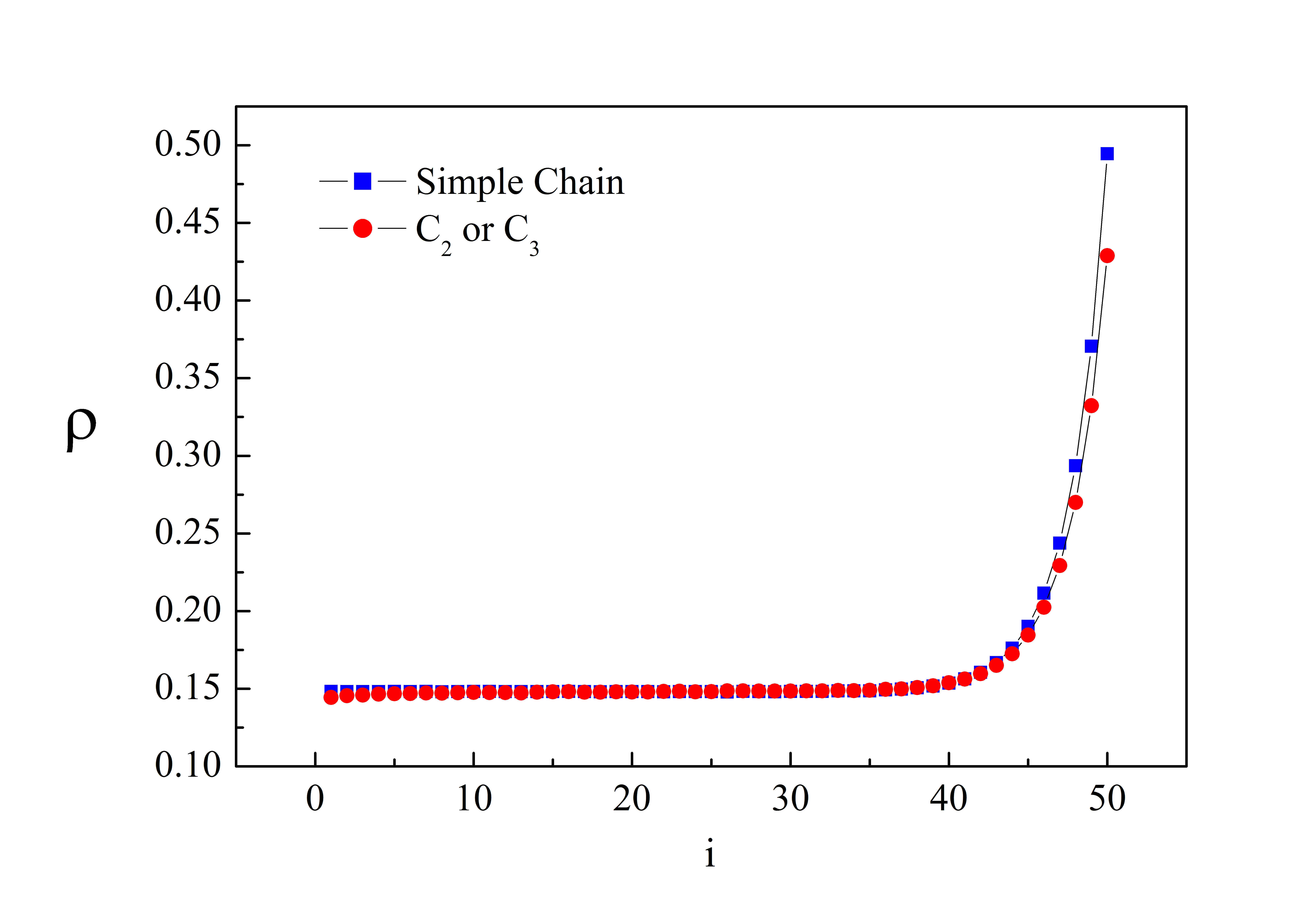}
   \caption{(Color online) Comparison of the simulation
   results for the particle density profile $\rho(i)$,
   $i=1, 2,\dots, 50$, in the LD phase of a chain
$C_{2,3}$ in the network (shown by solid (red) circles) and an
open simple chain (shown by solid (blue) squares) under the
estimated rates $\alpha = 0.148$ and $\beta = 0.255$.
\label{LDprofComp}}
  \end{center}
\end{figure}

Note that this value satisfies the conditions for
 the existence of a LD phase, but differs
significantly from the mean field estimates given in Table 1.  To
check the validity of the DW theory, we have performed computer
simulations of TASEP on a single chain with injection and ejection
rates $\alpha = 0.148$ and $\beta = 0.255$, respectively. The
corresponding density profiles are compared in Fig.
\ref{LDprofComp}.
One sees a qualitatively very close behavior of the density
profiles, although in the region of upward bending, close to the
right-hand end of the chains, the local density at the single
chain sites is considerably higher than the one at the
corresponding sites of the $C_{2,3}$ branches. At the right end
this difference reaches its maximum: $\rho^{2,3}(L_2) \simeq
0.429$, while for the simple chain $\rho^{SC}(L_2) \simeq 0.495$.
The latter value is not influenced significantly by finite-size
effects, because it coincides, within numerical accuracy, with the
value of the density in the thermodynamic limit for a simple chain
with the same boundary rates:
\begin{equation}
\lim_{L\rightarrow \infty} \rho_{LD}(L) =\alpha_d^*(1-\alpha)_d^*/\beta_d^* \simeq 0.494.
\label{rightend}
\end{equation}
We explain the above discrepancy by the presence of nearest neighbor correlations between
the last site of the chains $C_{2,3}$ and the first site of the tail chain $C_4$:
\begin{equation}
F^{(d,4)}_{nn} = \langle \tau_{L_d}^{(2,3)}\tau_1^{(4)}\rangle -
\langle \tau_{L_d}^{(2,3)}\rangle \langle \tau_1^{(4)}\rangle.
\label{corrd4}
\end{equation}
From the exact expression for the current $J^{(2,3)}$ through each of the chains $C_{2,3}$
it follows that
\begin{equation}
\rho_{L_d}^{(2,3)} = \frac{J^{(2,3)} + F^{(d,4)}_{nn}}{1 -\rho_1^{(4)}}.
\label{corrLd}
\end{equation}
Hence, by inserting the numerically evaluated $F^{(d,4)}_{nn} \simeq 0.048$ and $\rho_1^{(4)}
\simeq 0.590$, we obtain $\rho_{L_d}^{(2,3)} \simeq 0.424$ which is fairly close to the
simulations result $\rho^{2,3}(L_d) \simeq 0.429$.

Finally we note that although the DW theory gives an excellent description of the density
profile of chains in the LD phase, the maximum reached at the right-hand end of a long simple
chain is lower than the bulk density $\rho_+ = 1- \alpha_d^*$ of the high-density phase
that supports the same current for all $\alpha_d^* < \beta_d^*$:
\begin{equation}
\lim_{L\rightarrow \infty} \rho_{LD}(L) =\alpha_d^*(1-\alpha_d^*)/\beta_d^* <
(1-\alpha_d^*)= \rho_+ .
\label{roplus}
\end{equation}
Indeed, the interpretation $A=\rho_{+} - \rho_{-}$, given by the DW theory, yields a value
for the bulk density of the HD phase, $\rho_{+} = A + \rho_{-} \simeq 0.569$, which cannot
support the current $J^{(2,3)} = J^{max}/2$.

\subsubsection{Interpretation of the fit in the HD phase}

\begin{figure}
\begin{center}
  \includegraphics[width=100mm]{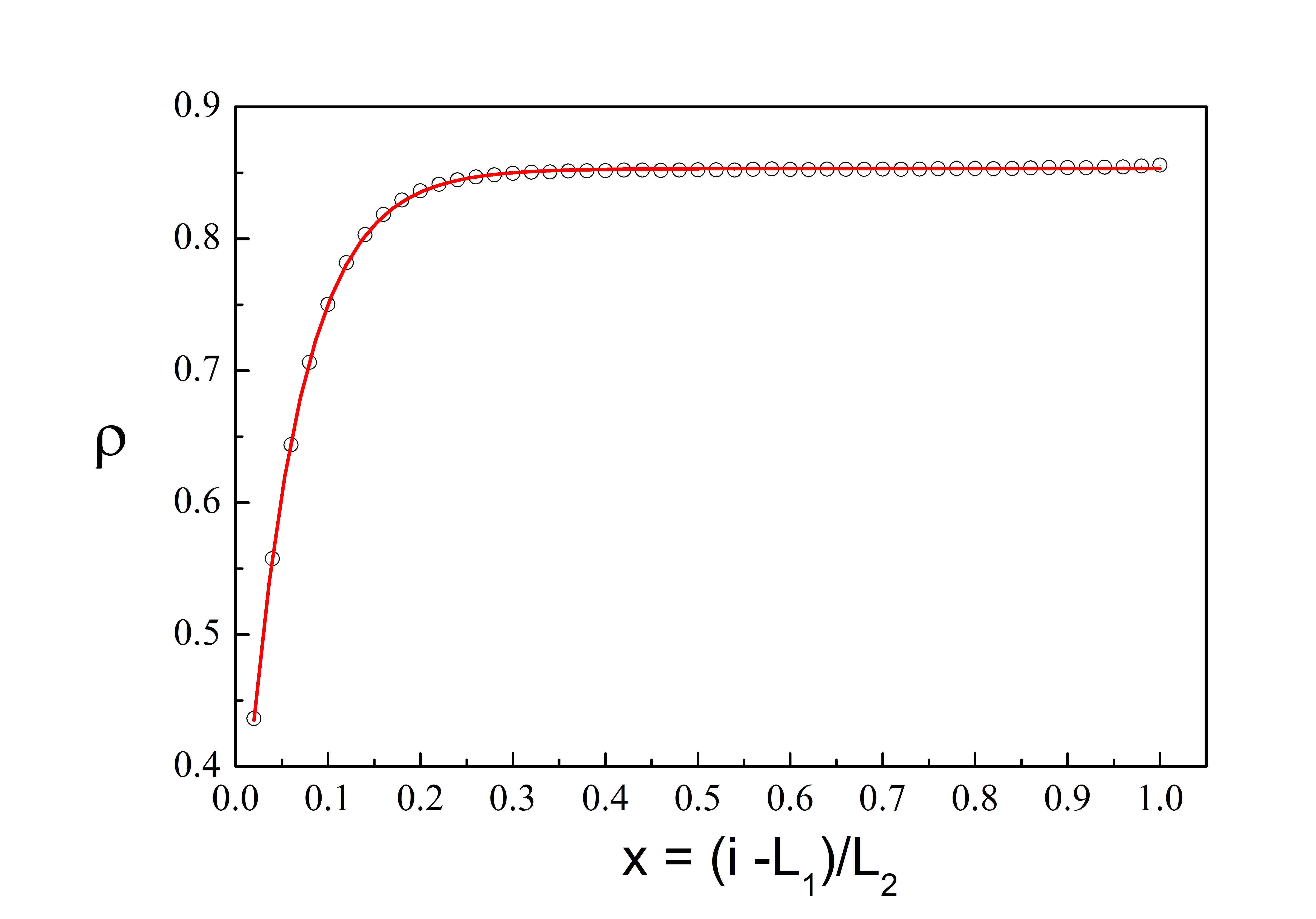}
   \caption{(Color online) The simulation results for the particle density
   profile $\rho(x)$, $x=(i-L_1)/L_2$, $i=L_1 +1,L_1 + 2,\dots, L_1 +L_2$, in the
HD phase of a chain $C_{2,3}$ in the network, are shown by
centered grey circles, the fit with the function (\ref{HDpro}) is
shown by solid (red) solid line. The quality of the fit is
characterized by the statistical criteria $\chi^2/DoF =9.134\times
10^{-7}$ and $R^2 = 0.99986$. \label{HDprof}}
  \end{center}
\end{figure}

The fit to the simulated local density profile by function (\ref{HDpro}) is shown in Fig. \ref{HDprof}.
Its excellent quality  yields reliable estimates of $\rho_{+} = 1-\beta_d^*
\simeq 0.853 >1/2$, $A = 0.594 \pm 0.002$ and $t = 0.0570 \pm 0.0002$, hence, $r = \exp(1/Lt) \simeq 1.42$.

From the relation for the bulk density in the HD phase, $\rho_{+}=1- \beta_d^*$, we deduce the value
$\beta_d^* \simeq 0.147$ for
the effective ejection rate. This gives a very good estimate of the
current $J^{(2,3)}=\beta_d^*(1-\beta_d^*) \simeq 0.1254$ through each of the chains $C_{2,3}$.

Next, following the DW theory, we solve the equation
\begin{equation}
\alpha_d^*(1-\alpha_d^*)= r \beta_d^*(1-\beta_d^*)=r J^{(2,3)}
\end{equation}
at  $r = 1.42$, to obtain
\begin{equation}
\alpha_d^* = (1/2)[1-\sqrt{1 -4r J^{(2,3)}}] \simeq 0.232.
\label{alphaHD}
\end{equation}
This value agrees fairly well with the mean field estimates given in Table 1.

\begin{figure}
\begin{center}
  \includegraphics[width=100mm]{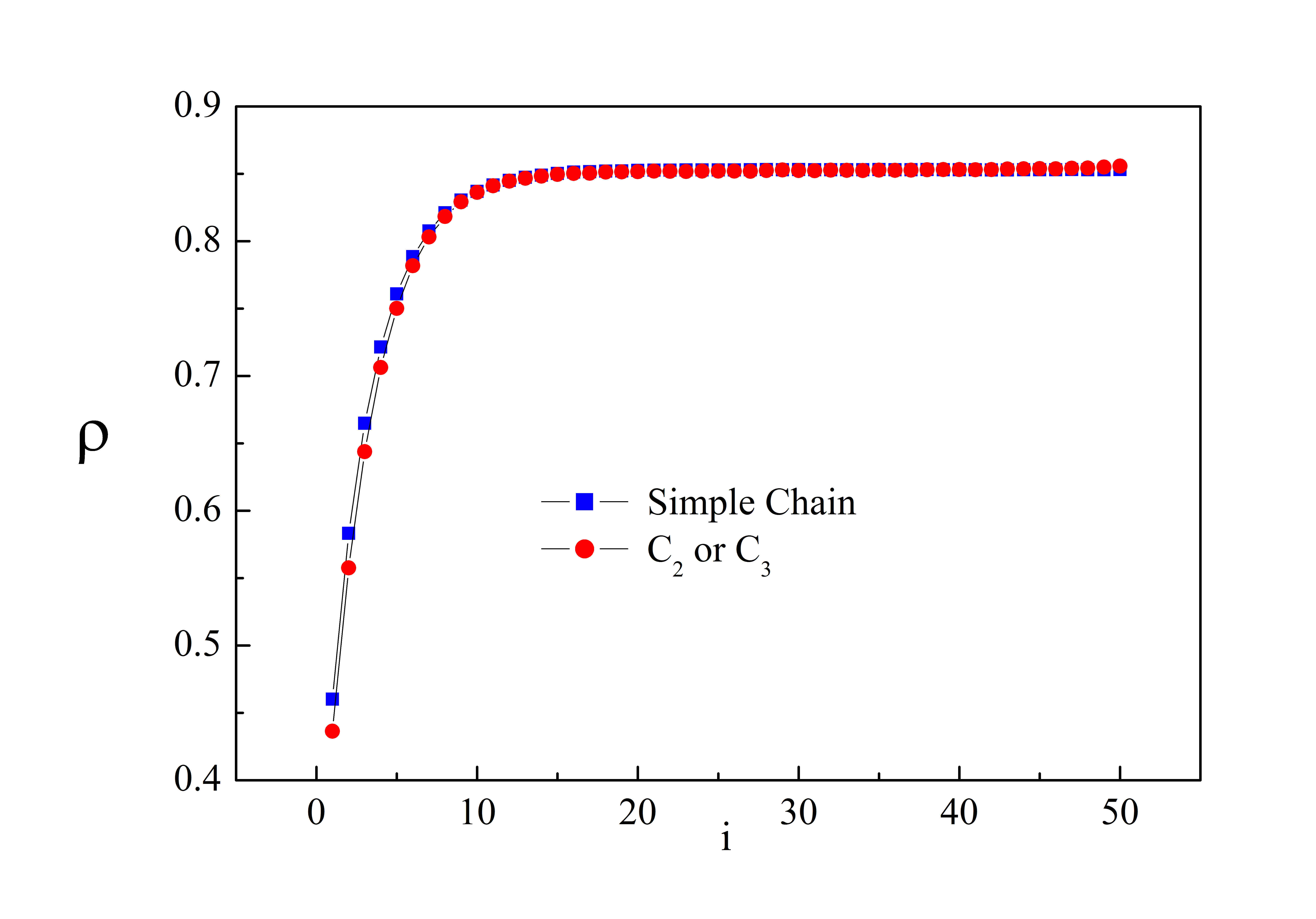}
    \caption{(Color online) Comparison of the simulation results for the particle density profile $\rho(i)$,
 $i=1, 2,\dots, 50$, in the HD phase of a chain
$C_{2,3}$ in the network (shown by solid (red) circles) and an
open simple chain (shown by solid (blue) squares) under the
estimated rates $\alpha = 0.232$ and $\beta =
0.147$.\label{HDprofComp}}
  \end{center}
\end{figure}

To check the consistency of the fit, from Eq. (\ref{HDprofF})
we obtain
\begin{equation}
\rho_- \simeq \rho_{HD}(1) =\rho_+ - A/r \simeq 0.435,
\end{equation}
which almost coincides
with the numerical value of $0.436$ for the density at the first site of
the chains $C_{2,3}$. On the other hand, the analytical expression for the
density profile in the thermodynamic limit leads to the estimate
\begin{equation}
\lim_{L\rightarrow \infty} \rho_{HD}(1)= 1 - \beta_d^*(1-\beta_d^*)/\alpha_d^*
\simeq 0.459,
\label{leftend}
\end{equation}
which coincides, within numerical accuracy, with the numerical value
$\rho^{SC}_{HD}(1)\simeq 0.460$ obtained for a simple chain under the same boundary
rates, see Fig. \ref{HDprofComp}, and
which is about 5 percent higher than the numerically obtained value $\rho_{HD}(1)=0.436$.
This small discrepancy is due to the small nearest-neighbor correlations between the
last site of the head chain and the first site of chains $C_{2,3}$:
\begin{equation}
F^{(1,d)}_{nn} = \langle \tau_{L_d}^{(1)}\tau_1^{(2,3)}\rangle -
\langle \tau_{L_d}^{(1)}\rangle \langle \tau_1^{(2,3)}\rangle
=\simeq 0.011. \label{corr1d}
\end{equation}

To check the validity of the DW predictions, we have simulated the density profiles
in a single chain with $L=50$, $\alpha = 0.232$ and $\beta = 0.147$.
A comparison of the results is given in Fig. \ref{HDprofComp}.

However, as in the previous case, the DW interpretation $A=\rho_+ -\rho_-$ leads to $\rho_- \simeq
0.259$, a value which does not support the current $J^{(2,3)}=J^{max}$ and, therefore cannot describe
a bulk LD phase.

\section{Discussion}

Non-equilibrium phenomena are much more often encountered and more
diverse in nature than the true equilibrium phenomena. The
development of our  understanding of physics far from equilibrium
is currently under way. In this respect  the study of simple
non-equilibrium models plays very important role. The asymmetric
simple exclusion process  is one of the simplest non-equilibrium
models of many-particle systems with particle conserving
stochastic dynamics and boundary induced phase transitions.

Here, we have reported results on a rather unexpected property of
the TASEP, defined on a simple network consisting of a single chain
with a double-chain insertion. In the symmetric case of
equal lengths of all the segments, and at equal
injection and ejection rates $\alpha =\beta >1/2$, the network has the
maximum-current (MC,CL,MC) phase structure. Then, a simple translation
of the double-chain section forward or backward along the single chain
induces a qualitative change in the local density profiles in the parallel
chains which is characteristic of a phase transition. This
``position-induced phase change'' is caused by the change of the
effective rates $\alpha^* $ and $\beta^*$ at the two junction
points of the network.

Using the continuity of the finite-size current $J_L$ one can
determine the effective rates $\alpha^*$ and $\beta^*$ at the two
ends of the loop. Our theoretical analysis, based on the results
for infinitely long chains, leads to values of $\alpha^*$ and $\beta^*$
which change with the double-section position and describe well
the phase changes obtained numerically in the profiles.
Note, for a finite-size system the effective rates depend on the
lengths of all the chains, $L_1$, $L_d$ and $L_4$.

Quite interestingly, upon increasing the number of parallel chains in the
inserted section, provided the parallel chains are in the LD phase, the total bulk
density of particles in them goes down to
to the value 1/4. Indeed, in the case of $n$ parallel equivalent
chains the current through each of them is $J^{max}/n = 1/(4n)$. The low
density phase that supports this current has a bulk density, compare
with Eq. (\ref{ropm}),
\begin{equation}
\rho_{-}(J^{max}/n) = \frac{1}{2}\left(1- \sqrt{1-n^{-1}}\right).
\label{Jn}
\end{equation}
Therefore,
\begin{equation}
\lim_{n\rightarrow \infty}n \rho_{-}(J^{max}/n) =
\lim_{n\rightarrow \infty}n\; \frac{1}{2}\left(1- \sqrt{1-n^{-1}}\right) = 1/4.
\label{nJLD}
\end{equation}
At that, the total bulk density in the $n$ parallel chains, when they are in the
HD phase, increases linearly with $n$
\begin{equation}
n \rho_{+}(J^{max}/n) = n - 1/4 +O(n^{-1}),
\quad \mbox{as} \quad n\rightarrow \infty.
\label{nJHD}
\end{equation}

Another novel observation, which we cannot explain completely, concerns
the validity of the DW description of the observed density profiles in the LD
and HD phases of chains
$C_{2,3}$. Apparently, see Figs. \ref{LDprof} and \ref{HDprof}, these profiles
are excellently fitted by a single exponential function, which implies the
existence of a well-defined localization length. However, the value of the
amplitude $A$ of the exponential function does not agree with the DW prediction
$A = \rho_+ - \rho_-$, where $\rho_+$ and $\rho_-$ are the bulk densities
of the HD and LD phases, respectively, which support the required current
$J^{max}/2$. This phenomenon cannot be attributed to correlations specific to
the network under consideration, because the same conclusions hold true for the
simple chains under the corresponding boundary rates.

\section*{Acknowledgment}
N.P. acknowledges a support by the Bulgarian Science Fund under
contract number D02-780/28.12.2012.

\end{document}